\documentclass[12pt]{iopart}

\usepackage{graphicx}

\begin{document}

\title[]{Thermal model predictions of hadron ratios at LHC}

\author{A Andronic$^1$, P Braun-Munzinger$^{1,2}$, J Stachel$^3$}
\address{$^1$ Gesellschaft f\"ur Schwerionenforschung, GSI, 
D-64291 Darmstadt, Germany}
\address{$^2$ Technical University Darmstadt, D-64283 Darmstadt, 
Germany}
\address{$^3$ Physikalisches Institut der Universit\"at Heidelberg,
D-69120 Heidelberg, Germany}

\begin{abstract}
We present predictions of the thermal model for hadron ratios in central 
Pb+Pb collisions at LHC.
\end{abstract}

Based on the latest analysis within the thermal model of the hadron yields 
in central nucleus-nucleus collisions \cite{Andronic:2005yp}, the
expected values at LHC for the chemical freeze-out temperature and 
baryochemical potential are $T$=161$\pm$4 MeV and 
$\mu_b$=0.8$^{+1.2}_{-0.6}$ MeV, respectively.
For these values, the thermal model predictions for hadron yield ratios 
in central Pb+Pb collisions at LHC are shown in Table~\ref{tab_aa_t1}.
We have assumed no contribution of weak decays to the yield of 
pions, kaons and protons.

\begin{table}[hbt]
\caption{Predictions of the thermal model for hadron ratios in central 
Pb+Pb collisions at LHC. The numbers in parantheses represent the error
in the last digit(s) of the calculated ratios.\\}
\label{tab_aa_t1}
\begin{tabular}{cccccc}
$\pi^-/\pi^+$ & $K^-/K^+$ & $\bar{p}/p$ & $\bar{\Lambda}/\Lambda$ & 
$\bar{\Xi}/\Xi$ & $\bar{\Omega}/\Omega$ \\ \br
1.001(0) & 0.993(4) & 0.948$^{-0.013}_{+0.008}$ & 0.997$^{-0.011}_{+0.004}$ & 
1.005$^{-0.007}_{+0.001}$ & 1.013(4) \\ \br \\
$p/\pi^+$ & $K^+/\pi^+$ & $K^-/\pi^-$ & $\Lambda/\pi^-$ & 
$\Xi^-/\pi^-$ & $\Omega^-/\pi^-$ \\ \br
0.074(6) & 0.180(0) & 0.179(1) & 0.040(4) & 0.0058(6) & 0.00101(15) \\ \br \\
\end{tabular}
\end{table}

The antiparticle/particle ratios are all very close to unity, with the
exception of the ratio $\bar{p}/p$, reflecting the expected small, 
but nonzero, $\mu_b$ value. The errors are determined by the errors of 
$\mu_b$ in case of antiparticle/particle ratios and by the errors of
$T$ for all other ratios. 

\begin{table}[hbt]
\caption{Predictions for the relative abundance of resonances at
chemical freeze-out.\\}
\label{tab_aa_t2}
\begin{tabular}{ccccccc}
$\phi/K^-$ & $K^{*0}/K^0_S$ & $\Delta^{++}/p$ & $\Sigma(1385)^+/\Lambda$ & 
$\Lambda^*/\Lambda$ & $\Xi(1530)^0/\Xi^-$ \\ \br
0.137(5) & 0.318(9) & 0.216(2) & 0.140(2) & 0.075(3) & 0.396(7)\\ \br
\end{tabular}
\end{table}

Assuming that the yield of resonances is fixed at chemical freeze-out,
we show in Table~\ref{tab_aa_t2} predictions for the relative yield of 
various resonance species.
We emphasize that the above hypothesis needs to be checked at LHC, 
in view of the data at RHIC \cite{Adams:2006yu}, which may indicate
rescattering and regeneration of resonances after chemical freeze-out.

\vspace{-.5cm}
\section*{References}
\bibliographystyle{unsrt}
\bibliography{therm}


\end{document}